# Real Space Observations of Magnesium Hydride Formation and Decomposition

T. Herranz,[†] K. F. McCarty,[‡] B. Santos,[†,◊] M. Monti,[†,◊] and J. de la Figuera[†,◊,1]

*[†]Instituto de Química-Física "Rocasolano", CSIC, Madrid 28006, Spain, [‡]Sandia National Laboratories, Livermore, California 94550, USA, [◊]CMAM, Universidad Autónoma de Madrid, 28049, Spain.*

Among the potential hydrides for H storage,[1] magnesium hydride ($MgH_2$) has been singled out due to magnesium's low cost, availability and lack of toxicity. Furthermore, large amounts of H, up to 7.6 Wt%, can be stored as $MgH_2$.[2] Well-known limitations are the thermodynamics requirements (up to 300ºC for H release) and slow kinetics. Possible solutions to address these problems are the use of small nanoparticles,[3] the alloying with transition metals,[4,5] or a combination of both approaches.[6] Surface-science studies, under ultra-high-vacuum (UHV) conditions, of well-controlled Mg surfaces interacting with H hold the promise to understand the limiting steps in magnesium hydriding/dehydriding. But such studies are scarce.[7] To avoid the problems involved in the use of Mg single crystals, thin films can be grown on several substrates with high perfection.[8,9] By this method, Chorkendorff *et al.* have performed hydriding studies of Mg(0001) films on Mo(111).[10] The characterization techniques employed were thermal desorption (TD) and x-ray photoelectron spectroscopy (XPS). The main findings confirm several ideas that had been put forward in less-controlled experiments, such as the lack of reactivity to molecular hydrogen[10], the deleterious effect of magnesium oxide on the hydride formation[11] and the possibility of using transition metals such as Pd[12] to dissociate $H_2$ and provide a source of atomic H. But to the best of our knowledge, no study has resolved the formation and decomposition of $MgH_2$ in real space. In this work, we study the growth and decomposition of thin layers of Mg and magnesium hydride on Ru(0001) using an in-situ technique that provides real-space, real-time observations of the formation of hydride islands.

The experiments have been performed using low-energy electron microscopy (LEEM).[13] This non-scanning technique allows for the fast (30 ms/image) imaging of a surface in UHV while depositing or desorbing material from the sample under study. Gases can be dosed while imaging at pressures up to $10^{-5}$ torr.[9] The changes in the surface structure during gas exposure can be followed in real time and the nucleation of new phases can be imaged without the limitations of other *in-situ* techniques (like X-ray diffraction or XPS) that only give average information. Also, selected-area low-energy electron diffraction (LEED) patterns can be acquired.[14]

Thin films of Mg on refractory metals are a classic system used to study electron confinement effects[15] and their influence on chemical reactivity.[9] Therefore, the thickness-dependent electronic properties can modify the behavior towards hydride formation, like in the case of nanoparticles. Unlike other substrates such as W(110), Mg growth on Ru(0001), our substrate, has been less studied.[8,16] Mg was deposited *in-situ* from a heated Mg rod. Hydrogen was dosed either molecularly by leaking from a reservoir of high-purity gas, or in atomic form using a hot W filament in a commercial cracker that provides up to 70% dissociated hydrogen.

Using LEEM, we have followed the growth of films up to 10 atomic layers (AL) for temperatures between 300 and 430 K. One of the advantages of this technique is the ability to grow and make measurements from film regions with a single, known thickness (see supporting information). As the first AL is completed, the density is already very similar to thicker films or bulk Mg, as detected by the change in the diffraction pattern. Thicker films were grown layer-by-layer. Exposure to molecular hydrogen with a total dose of over 700 L (22 minutes at $10^{-6}$ torr of total H) produces no detectable changes in the film as reported before.[10] In contrast, Figure 1 shows selected frames during exposure to the same dose of atomic hydrogen. Just after turning on the $H_2$ cracker (see Figure 1a), black islands start to nucleate and grow, the signature of a first-order phase transition.

Figure 1. LEEM images of a 10 AL Mg/Ru film exposed to atomic H at 305 K. The field of view (FOV) is 15 μm and the electron beam energy 1.25 eV. H dose: a) 90 L b) 250 L c) 850 L d) 1200 L

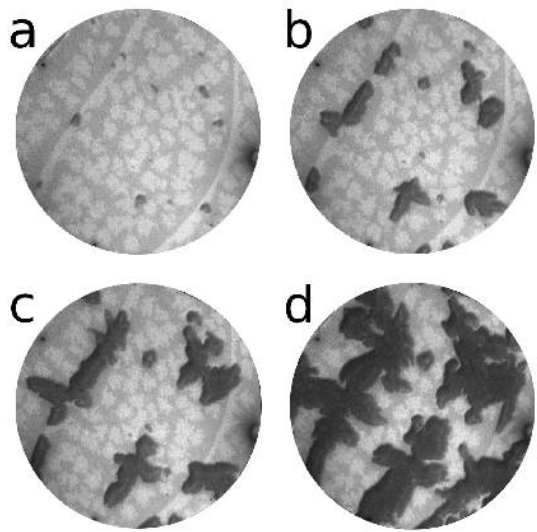

---

[1] *To whom correspondence should be addressed*

With increased exposure time, there is little further nucleation. Instead, the initially nucleated islands grow. This implies that nucleation is a rare event that requires a large supersaturation of H atoms on the Mg film, a supersaturation that is removed by the nucleation and growth. The lateral growth of the islands is initially fast but slows down markedly with time. Eventually they cover most of the field of view. The growth of black islands does not modify the distribution of the 10 AL Mg islands (the brightest regions in Figure 1), indicating that there is not significant Mg diffusion under atomic hydrogen exposure. The dark islands are stable in UHV at temperatures below 460 K.

The LEED patterns of the 4 AL Mg film in Figure 2a shows two sets of 6-fold diffraction spots, one set from the film and the other from the substrate. Attempts to obtain diffraction patterns from the black islands were frustrated because of the strong beam damage caused by electrons with kinetic energy higher than a few eV. So the LEED pattern of Figure 2b was taken at 15 eV and exposing the area to the electron beam for less than 1 second. The LEED patterns from black islands have sharp diffraction spots distinct from those of Ru(0001) and Mg(0001), showing that the black islands are crystalline. Different patterns were obtained from different islands, suggesting that they can have different orientations. LEEM images of the black islands during growth and decomposition were acquired using electrons of sufficiently low energy (0.6-1.3 eV) so that they only interact weakly with the surface (so-called mirror-mode microscopy).[13]

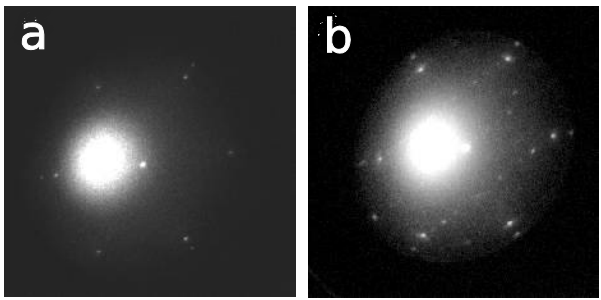

Figure 2. (a) LEED patterns from a 4 AL Mg film taken at 150 V (b) LEED pattern obtained from a black island attributed to MgH$_2$ (electron energy 15 V).

After exposing to hydrogen (either molecular or atomic), the films were heated in UHV and the surface was simultaneously imaged by LEEM and gas generation monitored by TD. The desorption of Mg exposed to molecular hydrogen is observed to take place layer by layer from 450 K. At 460 K, only the 1st AL of Mg remains on the Ru(0001) surface. This layer is more strongly bound to the substrate than to the Mg layers above and it desorbs at much higher temperatures, as reported by Over *et al.*[8] The evolution of the m/e=2 signal after molecular hydrogen exposure is a smooth function of temperature (see bottom curve in Figure 3). The very broad peaks are attributed to H desorption from the manipulator due to the non-ideal TD setup (see supporting information), and correlate well with the m/e=28 signal (CO).

The behavior of the films exposed to atomic hydrogen, where the growth of black islands was observed, is markedly different. In the LEEM images no changes were observed up to temperatures around 450 K. From this temperature desorption of the Mg layers started. This desorption occurs layer-by-layer not observing Ostwald ripening prior to desorption. As shown in Figure 3 (in a film originally composed of 8 Mg AL with some 9 AL islands) when the temperature reaches 463 K the dark islands start to disappear. By 479 K the islands are gone, leaving only a "shadow" of where they were.

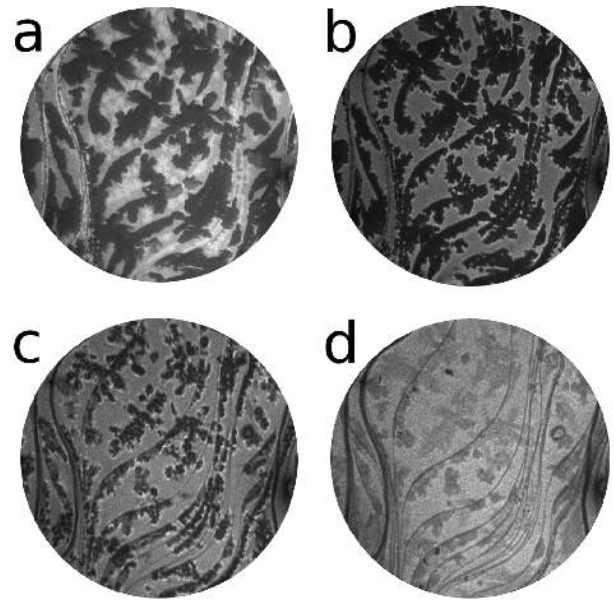

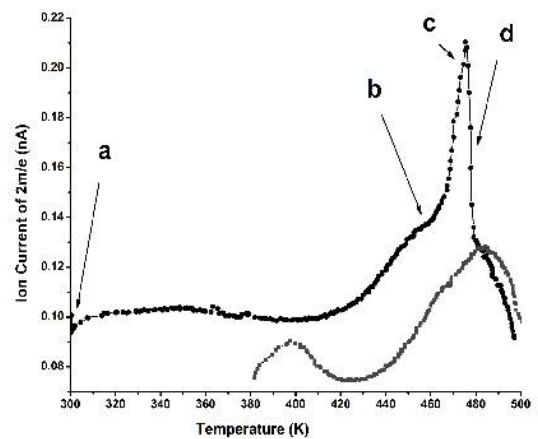

Figure 3. Upper: LEEM images during thermal decomposition of hydride. The FOV is 15 μm. Plot: TD signal of m/e=2. The lower curve corresponds a 9 AL Mg/Ru film exposed to H$_2$; the upper curve to a similar film exposed to H. Labeled temp. correspond to the images a-d.

Correlated with the disappearance of the black islands, there is a sudden spike in the m/e=2 TD signal, as seen in the upper curve in Figure 3. No spikes were detected in any other signal (m/e of 18, 24, 28 and 32 amu were monitored). This implies that the hydrogen release and the decomposition of the black islands produced during exposure to H are linked. Consequently we identify the black islands as magnesium hydride. The temperature at which the spike in the m/e=2 signal appears is in agreement with the one observed by Chorkendorff *et al.* for the decomposition of Mg hydride thin films (less than 10 AL).[10] In agreement with the relative stability of magnesium and its hydride, the decomposition temperature of the hydride is by about 10 K higher than the sublimation temperature of Mg itself.[10]

We have explored the dependence of the magnesium hydride islands as a function of initial Mg thickness and H dose. As shown in Figure 4, the decomposition temperature is reduced by ~10 K when the Mg film decreases from 9 to 4 AL. For comparison, Figure 4S shows that the variation peak desorption temperatures of repeated experiments is rather smaller, ±2 K. An obvious question is why the starting thickness of the Mg layers should have any influence on the MgH$_2$ stability. This is specially puzzling because by the time the hydride starts to decompose there is hardly any magnesium left on the Ru(0001) substrate, except for the initial Mg layer. A likely explanation is that the film thickness influences the thickness of the hydride grown on top: thicker films can accommodate thicker hydride islands. The depth of the hydride islands cannot be determined from the LEEM images. However, the slowdown of the lateral growth under atomic H exposure could imply that the growth of the hydride continues into the Mg film forming 3D islands. The decomposition temperature of hydride islands grown for a film with the same thickness, but with a dose of 100 L (~1/10 of the dose employed in Figure 4) is reduced by 20 K, as detected by LEEM. This result can be reconciled with the data shown in Figure 4 if we assume that the hydride islands grew during the initial stages mostly in-plane (bi-dimensional shape), and only later (if the film is thick enough) three-dimensionally.

In summary, we have followed in real space and real time the growth and thermal desorption of Mg hydride on Ru(0001) by means of simultaneous LEEM and TD. The disappearance of the hydride coincides with a spike in the m/e=2 signal in the TD. The temperature of H$_2$ evolution in our system is similar to the one reported in previous works dealing with nanoparticles[3] and thin films[10,11] and lower than the one reported for bulk MgH$_2$.[1] We have studied the effect of H dose and Mg film thickness on the growth kinetics and on the thermal desorption of the hydride. An increase in the decomposition temperature is observed with thicker original Mg films. This is attributed to the more 3D character of the hydride grown and explains why Mg nanoparticles dehydride at lower temperatures than bulk Mg[3]. This real-space nanometric study therefore provides a way to explore the kinetic limitations on MgH$_2$ formation and decomposition.

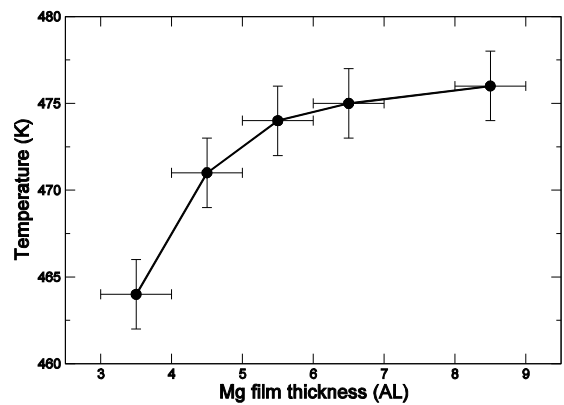

Figure 4. Dependence of the peak of the 2 amu signal as a function of the Mg film thickness before exposing to H. Hor. error bars: presence of islands with ±1 AL in the film. Vert. error bars: ±2 K (variance in the temp of the spike in H$_2$ evolution observed in control experiments).

Acknowledgment: This research was supported by the Office of Basic Energy Sciences, Division of Materials and Engineering Sciences, U.S. DOE under Contract No. DE-AC04-94AL85000 and the Spanish Ministry of Innovation and Science under Project No. MAT2006-13149-C02-02.

Supporting Information Available. This information is available free of charge via the Internet at http://pubs.acs.org.

### References

(1) Grochala, W.; Edwards, P. P. Chem. Rev. 2004, 104, 1283-1316.

(2) Thomas, G.; Sandrock, G. DOE/SNL Hydride database, http://hydpark.ca.sandia.gov.

(3) Aguey-Zinsou, K.F., Ares-Fernández, J.R., Chem. Mater. 2008, 20, 376-378.

(4) Martin, M.; Gommel, C.; Borkhart, C.; Fromm, E. J. Alloys Comp. 1996, 238, 193-201.

(5) Chen, J.; Yang, H.; Xia, Y.; Kuriyama, N.; Xu, Q.; Sakai, T. Chem. Mater. 2002, 14, 2834-2836.

(6) Liang, G.; Huot, J.; Boily, S.; Neste, A. V.; Schulz, R. J. Alloys Comp. 1999, 292, 247-252.

(7) Sprunger, P.; Plummer, E. Chem. Phys. Lett. 1991, 187, 559-564.

(8) Over, H.; Hertel, T.; Bludau, H.; Panz, S.; Ertl, G. Phys. Rev. B 1993, 48, 5572-5578.

(9) Aballe, L.; Barinov, A.; Locatelli, A.; Heun, S.; Kiskinova, M. Phys. Rev. Lett. 2004, 93, 196103.

(10) Ostenfeld, C. W.; Davies, J. C.; Vegge, T.; Chorkendorff, I. Surf. Sci. 2005, 584, 17-26.

(11) Ostenfeld, C. W.; Chorkendorff, I. Surf. Sci. 2006, 600, 1363-1368.

(12) Johansson, M.; Ostenfeld, C. W.; Chorkendorff, I. Phys. Rev. B 2006, 74, 193408.

(13) Bauer, E. Rep. Progr. Phys. 1994, 57, 895-938.

(14) de la Figuera, J.; Puerta, J.; Cerda, J.; Gabaly, F. E.; McCarty, K. Surf. Sci. 2006, 600, L105-L109.

(15) Schiller, F.; Heber, M.; Servedio, V. D. P.; Laubschat, C. Phys. Rev. B 2004, 70, 125106.

(16) Schwegmann, S.; Over, H.; Gierer, M.; Ertl, G. Phys. Rev. B 1996, 53, 11164.